\documentclass[12pt, a4paper]{article}
\usepackage[font=footnotesize]{caption}
\usepackage[utf8]{inputenc}
\usepackage{graphicx}
\usepackage{authblk}
\usepackage{booktabs}
\usepackage{verbatim}
\usepackage{url}
\usepackage{caption}
\usepackage{subcaption}
\usepackage{indentfirst}
\usepackage{url}
\usepackage{amsmath}
\author{Alexandre Benatti$^1$, Henrique F. de Arruda$^{1,2}$ \\ and Luciano da F. Costa$^1$}

\date{%
    $^1$S\~ao Carlos Institute of Physics,
    University of S\~ao Paulo, S\~ao Carlos, SP, Brazil.\\%
    $^2$CENTAI Institute, Turin, Italy. \\[2ex]%
    6th Mar. 2024
}

\begin{document}

\title{Neuromorphic Networks as Revealed by Features Similarity}

\maketitle
\begin{abstract}
The study of neuronal morphology is important not only for its potential relationship with neuronal dynamics, but also as a means to classify diverse types of cells and compare than among species, organs, and conditions. In the present work, we approach this interesting problem by using the concept of coincidence similarity, as well as a respectively derived method for mapping datasets into networks. The coincidence similarity has been found to allow some specific interesting properties which have allowed enhanced performance (selectivity and sensitivity) concerning several pattern recognition tasks. Several combinations of 20 morphological features were considered, and the respective networks were obtained by maximizing the literal modularity (in supervised manner) respectively to the involved parameters. Well-separated groups were obtained that provide a rich representation of the main similarity interrelationships between the 735 considered neuronal cells. A sequence of network configurations illustrating the progressive merging between cells and groups was also obtained by varying one of the coincidence parameters.
\end{abstract}

\maketitle

\section{\label{sec:introduction}Introduction}

The natural world can be understood as a system composed of matter, energy, and information interconnected throughout time and space. So is the brain of living beings, where neuronal cells process and transmit information across an intricate network of synaptic interconnections.  

Though several complex systems --- such as the Internet, energy transmission, and transportation --- involve intricate interconnecting networks, the brain possesses a specific interesting property in which the \emph{shape} of its basic processing components, the neurons, can influence the respective topology and dynamics (e.g.~\cite{Ascoli, da2021multiset, yuste2015neuron, cajal1989recollections, friedman2020measurements, grueber2005development, da2002shape}). One direct manner in which this takes place is that neurons with more intricate dendritic arborizations tend to cover more effectively the surrounding 3D space, therefore increasing the chances of receiving axonal projections from other neurons, which are frequently guided by fields emanating from the dendritic arborizations.

Since the pioneering works by Santiago Ramon-y-Cajal (e.g.~\cite{ramon1894croonian, ramon1906structure, y1928degeneration}) the importance of neuronal shape has drawn continued interest, having motivated a large number of related works (e.g.~\cite{Ascoli, da2021multiset, yuste2015neuron, cajal1989recollections, friedman2020measurements, grueber2005development, da2002shape,donohue2011automated, langhammer2010automated, kunst2019cellular, linden2014lfpy, vanwalleghem2018integrative, gouwens2018systematic}).

Among the several interesting resources that have been made available on the Internet, the \emph{Neuromorpho} database (NeuroMorpho.Org, ~\cite{ascoli2007neuromorpho,akram2018open}) incorporates an impressive number of 3D neuronal cell reconstructions that are also accompanied by a set of several measurements or \emph{features}, such as the number of bifurcations, the number of terminal tips, the order of the branch concerning soma, and other \emph{features}. 

NeuroMorpho.org, therefore, paves the way for a wide range of possible studies on neuronal morphology, such as comparing the similarity between cells from diverse organisms, organs, types, etc. Indeed, much of the morphological classification of neuronal cells has been performed based on comparisons between a representative set of respective measurements. This type of research is of particular importance as it provides the background for establishing possible relationship between neuronal morphology and dynamics (e.g.~\cite{Ascoli, da2021multiset, yuste2015neuron, cajal1989recollections, friedman2020measurements, grueber2005development, da2002shape, de2015framework, linden2014lfpy}).

Comparisons between neuronal cell morphology have frequently relied on taking distances between the respective features. Similarity-based approaches have been relatively less frequent, often adopting indices such as the cosine similarity and the Pearson correlation coefficient. However, these approaches have some specific characteristics, including the fact that the cosine similarity does not take into account the magnitude of the feature vectors and the potential susceptibility of the Pearson correlation to outliers, as well as its tendency to yield biased results when few samples are available (e.g.~\cite{kim2015instability}).

\begin{figure}[!ht]
  \centering
    \includegraphics[width=.22\textwidth]{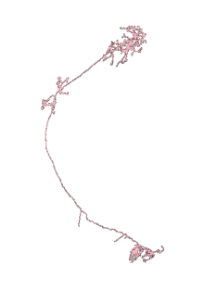}
    \includegraphics[width=.22\textwidth]{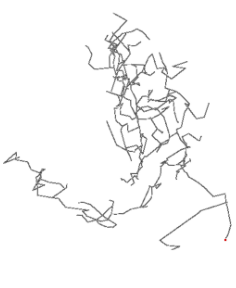}
    \includegraphics[width=.22\textwidth]{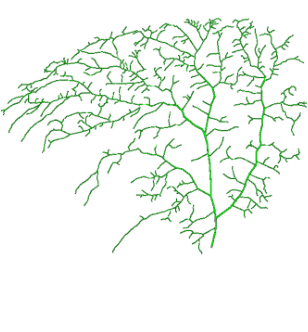} 
    \includegraphics[width=.22\textwidth]{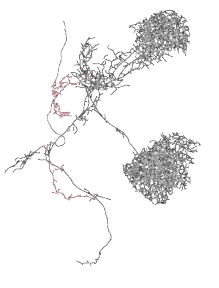} 
  \caption {Illustration of four neurons considered in the present work~\cite{zheng2018complete,shih2015connectomics}. The impressively diversified morphology of biological neurons can influence neuronal interconnectivity and dynamics.}
  \label{fig:Neurons}
\end{figure}

Introduced more recently~\cite{CostaCCompl, da2021further}, the \emph{coincidence similarity} relies on an extension of the Jaccard index to real values as well as a complementation of the similarity quantification by taking into account the relative interiority between non-zero feature vectors, given that the Jaccard similarity has been shown not to be able to reflect the latter property~\cite{da2021further}. The coincidence similarity has some important intrinsic features that contribute to its enhanced selectivity and stability~\cite{da2021multiset}. These properties have paved the way for several respective applications (e.g.~\cite{domingues2022city, benatti2022retrieving, dos2022enzyme, da2021multiset, da2021segmentation, costa2022similarity, da2021elementary}), including the translation of datasets, in which the elements are described by features, into respective networks~\cite{CostaCCompl}. This approach has been called, for simplicity's sake, \emph{coincidence similarity methodology}. Basically, after possible eventual normalization of the involved features, the coincidence similarity between each data element is quantified and associated with respective links in the resulting network.

The coincidence similarity operation can be controlled by three parameters, namely a global threshold $T$, an exponent $D$, as well as a parameter $\alpha$ controlling the relative contribution of features with the same or opposite signs to the overall similarity result. The flexibility allowed by these parameters has been effectively employed as a means of optimizing the properties of the networks derived from datasets by using the coincidence method (e.g.~\cite{CostaCCompl}). In addition, as the parameter $\alpha$ is successively increased, the network becomes more and more interconnected while preserving the previous connections. Thus, by starting with a small value of $\alpha$ and then increasing it progressively, it becomes possible to derive important insights into how the types of patterns are interrelated.

In the present work, we approach the important topic of comparing and organizing neuronal morphology according to their mutual similarity to be gauged in terms of the real-valued coincidence index~\cite{da2021further, costa2021similarity}. The main point characterizing this approach
relates to the enhanced selectivity and sensitivity while comparing non-zero vectors of morphological measurements accounted by the coincidence similarity, being therefore specific to applications requiring these properties. In addition, it should be taken into account that the coincidence similarity, as the Jaccard index from which it derives, are not intrinsically invariant to operations such as rotation, though this property would anyway require the considered features to correspond to an orthonormal system of coordinates.

A total of 20 morphological measurements were obtained from NeuroMorpho.org with respect to the 8 cell types of the \emph{Drosophila melanogaster}~\cite{zheng2018complete,boergens2018full,	cuntz2013preserving,heckscher2015even,takagi2017divergent,	zwart2016selective,eichler2017complete,schlegel2016synaptic,lo2015quantification,chiang2011three,shih2015connectomics}. The main motivation is to explore the special selectivity of the coincidence method for deriving networks respectively to the rich morphological data provided by NeuroMorpho.org while exploring the above-mentioned possibilities allowed by this methodology. More specifically, we aim at revealing the interrelationship of a fixed set of neuronal cells as quantified by the coincidence similarity at the highest respective literal modularity~\cite{da2022literal} so as to impose the most selective and strict comparison, which is performed in supervised manner.

In addition, we also aim at studying the effect of each of the adopted morphological features respectively on the interconnection and literal modularity of the obtained networks of neuronal types. In particular, we obtain the networks after removing each of the adopted features while quantifying the respective impact on the results. A network can then be constructed based on the similarity between the networks obtained by the different combinations of features considered.

Several interesting results have been obtained and discussed in this work. From the perspective of literal modularity supervised optimization, it has been verified that the exponent parameter ($D$) can strongly influence the selectivity and sensitivity of the parameter space where the larger literal modularity values are located. Regarding the study of the features influence on the obtained results, performed based on coincidence similarity between the maximum modularity networks obtained for several feature combinations, it was verified that each of the features has a relatively moderate effect on the results, yielding networks with mostly comparable modularity. The neuromorphic networks obtained by the coincidence method resulted not only highly modular, in which not only compact clusters have been obtained respectively to the considered neuronal types, but also provided a detailed representation of the similarity interrelationships between the several neurons from the same and different groups. The adopted methodology also allowed the identification of outlier cells.

Each of the identified groups has also been visualized separately, allowing a better appreciation of their inner interconnections that are revealed by the coincidence similarity method. Then, by incorporating the parameter ($\alpha$), coincidence similarity networks were obtained which effectively highlight the successive merging of cells and groups, providing additional insights about their specific interrelationships.

This article is organized as follows. It starts by presenting the several resources, concepts, and methods adopted in this work, including the dataset, the considered morphological measurements, the adopted literal modularity, the concept of multiset coincidence similarity, and the derived method for translating datasets into networks, as well as the methodology for a systematic study of the effect of the feature combinations on the obtained results. This was followed by the presentation and discussion of the results respectively to the modularity maximization, the consideration of several features combinations, and the resulting maximum modularity coincidence network revealing in a comprehensive and detailed manner the interrelationships between considered neuronal reconstructions from the point of view of the similarity between their morphological characteristics.

\section{Methodology}\label{sec:methods}

In this section, we describe the considered dataset and respective neuromorphic measurements, the adopted literal modularity methodology, the real-valued coincidence similarity, its application to translated datasets to complex networks, as well as a description of a method for studying the influence of feature selection on the obtained results.

\subsection{Dataset and measurements}

In order to illustrate the derivation of networks defined by the similarity between the properties of neuronal cells, we selected eight types of neurons of \emph{Drosophila melanogaster} from the NeuroMorpho.org database~\cite{ascoli2007neuromorpho,akram2018open}. Primarily aimed at sharing detailed neuronal reconstructions, the NeuroMorpho.org resource contains over 150.000 reconstructions from more than 800 laboratories. It was started and has since then been maintained by the Computational Neuroanatomy Group at the Krasnow Institute for Advanced Study, George Mason University~\cite{ascoli2007neuromorpho}.

A total of 20 measures have been taken as features of the reconstructed neuronal cells in our approach. These measurements\footnote{http://cng.gmu.edu:8080/Lm/help/index.htm}, see also~\cite{scorcioni2008measure}, were chosen as corresponding to the entries that were available for the selected cells. In other words, a measurements that were not available from some of those cells were not included in our analysis. In the following, we summarize the 20 adopted measurements.

\begin{itemize}
\footnotesize

\setlength\itemsep{0.1em}
    \item \emph{Surface}: The surface area of the compartment, calculated at each tracing point, with unit $\mu m^2$;
    \item \emph{Volume}: The volume of the compartment, calculated at each tracing point, with unit $\mu m^3$;
    \item \emph{N\_stems}: The number of stems attached to the \emph{Soma} in count units;
    \item \emph{N\_bifs}: The number of bifurcations in count units;
    \item \emph{N\_branch}: The total number of neuron branches in count units;
    \item \emph{Width}: Length of the entire neuron arbor along the most distributed
    PCA projection, with unit $\mu m$;
    \item \emph{Height}: Length of the entire neuron arbor along the second most distributed
    PCA projection, with unit $\mu m$;
    \item \emph{Length} of the entire neuron arbor along the third most distributed
    PCA projection, with unit $\mu m$;
    \item \emph{Diameter}: The total sum of the diameter of each compartment of the neuron, with the unit $\mu m$;
    \item \emph{EucDistance}: The sum of the straight line distance of each compartment from the \emph{Soma} calculated for each compartment, with the unit $\mu m$;
    \item \emph{PathDistance}: The sum of lengths of each compartment from the \emph{Soma} point, with the unit $\mu m$;
    \item \emph{Branch\_Order}: The order of the branch, in count units, where \emph{Soma} has order=$0$;
    \item \emph{Contraction}:  The ratio between \emph{EucDistance} and \emph{PathDistance};
    \item \emph{Fragmentation}: The sum of the number of compartments that constitute a branch between two bifurcation points, calculated at each bifurcation point;
    \item \emph{Partition\_asymmetry}: $\frac{\vert n_1 - n_2\vert}{n_1+n_2-2}$, where $n_1$ is the number of tips on the left of a bifurcation and $n_2$ is the number of tips on the right, calculated at each bifurcation point;
    \item \emph{Pk\_classic}: The ratio between the sum of the diameter of two daughter branches  by the diameter of the bifurcating parent, calculated at each bifurcation point;
    \item \emph{Bif\_ampl\_local}: The angle between branches at a bifurcation, with units in degree;
    \item \emph{Fractal\_Dim}: The slope of the linear fit of regression line obtained from the log-log plot of \emph{PathDistance} vs \emph{EucDistance}, for each branch;
    \item \emph{Bif\_ampl\_remote}: The angle between two terminal points of a bifurcation, with unit degree;
    \item \emph{Length}: The length between the two endpoints of a compartment, with the units in $\mu m$.
\end{itemize}

We chose neurons from the \emph{Drosophila melanogaster} because of the special importance of this species as a model organism in several biological areas, including neuroscience (e.g.~\cite{cardona2010identifying, schnaitmann2018color, scheffer2020connectome}). The eight types of neurons~\cite{zheng2018complete,boergens2018full,cuntz2013preserving,heckscher2015even,takagi2017divergent,zwart2016selective,eichler2017complete,schlegel2016synaptic,lo2015quantification,chiang2011three,shih2015connectomics}, summarized in Table~\ref{tab:types}, were selected so as to provide a diversity of cell types and morphologies. The selected eight types are presented in Table~\ref{tab:features1} and Table~\ref{tab:features2}, which show the number of cells considered, as well as the mean and average values of the respectively adopted measures.

\begin{table}[]
\centering
\caption{The eight types of neurons considered in the present work, as well as the
the respective number of cells that have been taken into account.}\label{tab:types}
\begin{tabular}{|c|c|c|}
\hline
Type & Name                                                                                                                & \begin{tabular}[c]{@{}c@{}}Number\\ of  Neurons\end{tabular} \\ \hline
1    & \begin{tabular}[c]{@{}c@{}}Antennal \\ lobe-uniglomerular\\ projection~\cite{zheng2018complete}\end{tabular}                                 & 122                                                                                  \\ \hline
2    & \begin{tabular}[c]{@{}c@{}}Borst-optic\\ Lobe-tangential~\cite{boergens2018full,cuntz2013preserving}\end{tabular}                                               & 34                                                                                   \\ \hline
3    & Cardona-Motoneuron~\cite{heckscher2015even,takagi2017divergent,zwart2016selective}                                                                                                  & 80                                                                                   \\ \hline
4    & \begin{tabular}[c]{@{}c@{}}Protocerebrum-Kenyon cell~\cite{eichler2017complete}\end{tabular}                                                 & 215                                                                                  \\ \hline
5    & \begin{tabular}[c]{@{}c@{}}Subesophageal \\ zone-(SEZ)-principal cell~\cite{schlegel2016synaptic}\end{tabular}                                & 37                                                                                   \\ \hline
6    & \begin{tabular}[c]{@{}c@{}}Chen B-peripheral nervous \\ system-Multidendritic-dendritic\\ arborization~\cite{lo2015quantification}\end{tabular} & 99                                                                                   \\ \hline
7    & \begin{tabular}[c]{@{}c@{}}Chiang-adult\\ centralcomplex-cholinergic~\cite{chiang2011three}\end{tabular}                                   & 104                                                                                  \\ \hline
8    & \begin{tabular}[c]{@{}c@{}}Adult subesophageal\\ zone-GABAergic~\cite{shih2015connectomics,chiang2011three}\end{tabular}                                        & 44                                                                                   \\ \hline
\end{tabular}
\end{table}

\begin{table}[]
\caption{The average $\pm$ standard deviation of each of the 20 morphological measurements for the
neuronal types 1 to 4.}\label{tab:features1}
\scriptsize

\begin{tabular}{l|llll}
                              & \textbf{Type 1}                & \textbf{Type 2}                & \textbf{Type 3}               & \textbf{Type 4}               \\ \hline
\textbf{Surface}              & 1060.0 $\pm$ 470.0  & 2900.0 $\pm$ 2400.0 & 162.0 $\pm$ 58.0   & 200.0 $\pm$ 110.0  \\
\textbf{Volume}               & 66.0 $\pm$ 29.0     & 1110.0 $\pm$ 980.0  & 25.0 $\pm$ 89.0    & 12.2 $\pm$ 6.6     \\
\textbf{N\_stems}             & 1.98 $\pm$ 0.13     & 1.21 $\pm$ 0.4      & 1.29 $\pm$ 0.45    & 1.009 $\pm$ 0.096  \\
\textbf{N\_bifs}              & 230.0 $\pm$ 220.0   & 390.0 $\pm$ 240.0   & 98.0 $\pm$ 53.0    & 101.0 $\pm$ 68.0   \\
\textbf{N\_branch}            & 470.0 $\pm$ 430.0   & 770.0 $\pm$ 490.0   & 200.0 $\pm$ 110.0  & 200.0 $\pm$ 140.0  \\
\textbf{Width}                & 48.0 $\pm$ 17.0     & 79.0 $\pm$ 31.0     & 10.2 $\pm$ 4.8     & 17.6 $\pm$ 8.3     \\
\textbf{Height}               & 209.0 $\pm$ 30.0    & 99.0 $\pm$ 33.0     & 32.7 $\pm$ 9.9     & 35.0 $\pm$ 11.0    \\
\textbf{Depth}                & 120.0 $\pm$ 25.0    & 50.0 $\pm$ 48.0     & 18.9 $\pm$ 9.7     & 14.7 $\pm$ 7.0     \\
\textbf{Diameter}             & 0.25 $\pm$ 0.0      & 0.28 $\pm$ 0.22     & 0.264 $\pm$ 0.031  & 0.25 $\pm$ 0.0     \\
\textbf{EucDistance}          & 168.0 $\pm$ 26.0    & 103.0 $\pm$ 21.0    & 34.9 $\pm$ 8.0     & 40.0 $\pm$ 11.0    \\
\textbf{PathDistance}         & 320.0 $\pm$ 120.0   & 167.0 $\pm$ 59.0    & 54.0 $\pm$ 13.0    & 83.0 $\pm$ 28.0    \\
\textbf{Branch\_Order}        & 30.0 $\pm$ 13.0     & 37.0 $\pm$ 12.0     & 27.0 $\pm$ 10.0    & 26.0 $\pm$ 13.0    \\
\textbf{Contraction}          & 0.856 $\pm$ 0.024   & 0.918 $\pm$ 0.038   & 0.894 $\pm$ 0.017  & 0.871 $\pm$ 0.023  \\
\textbf{Fragmentation}        & 4200.0 $\pm$ 1900.0 & 3100.0 $\pm$ 1300.0 & 1060.0 $\pm$ 360.0 & 1440.0 $\pm$ 780.0 \\
\textbf{Partition\_asymmetry} & 0.632 $\pm$ 0.037   & 0.623 $\pm$ 0.045   & 0.663 $\pm$ 0.093  & 0.64 $\pm$ 0.11    \\
\textbf{Pk\_classic}          & 2.0 $\pm$ 0.0       & 1.7 $\pm$ 0.25      & 2.02 $\pm$ 0.2     & 2.0 $\pm$ 0.0      \\
\textbf{Bif\_ampl\_local}     & 96.9 $\pm$ 4.5      & 83.0 $\pm$ 13.0     & 88.0 $\pm$ 13.0    & 92.2 $\pm$ 8.2     \\
\textbf{Fractal\_Dim}         & 1.171 $\pm$ 0.03    & 1.06 $\pm$ 0.022    & 1.138 $\pm$ 0.034  & 1.13 $\pm$ 0.032   \\
\textbf{Bif\_ampl\_remote}    & 93.5 $\pm$ 3.8      & 79.2 $\pm$ 7.3      & 89.0 $\pm$ 8.7     & 88.0 $\pm$ 12.0    \\
\textbf{Length}               & 1350.0 $\pm$ 600.0  & 2090.0 $\pm$ 790.0  & 202.0 $\pm$ 68.0   & 250.0 $\pm$ 140.0 
\end{tabular}

\end{table}

\begin{table}[]
\caption{The average$\pm$ standard deviation of each of the 20 morphological measurements for the
neuronal types 5 to 8.}\label{tab:features2}
\scriptsize
\begin{tabular}{l|llll}
                              & \textbf{Type 5}               & \textbf{Type 6}                 & \textbf{Type 7}                & \textbf{Type 8}                \\ \hline
\textbf{Surface}              & 330.0 $\pm$ 110.0  & 20100.0 $\pm$ 8300.0 & 8900.0 $\pm$ 7600.0 & 7200.0 $\pm$ 9300.0 \\
\textbf{Volume}               & 30.0 $\pm$ 37.0    & 5000.0 $\pm$ 2100.0  & 2200.0 $\pm$ 1900.0 & 1800.0 $\pm$ 2300.0 \\
\textbf{N\_stems}             & 1.32 $\pm$ 0.47    & 1.79 $\pm$ 0.41      & 1.0 $\pm$ 0.0       & 1.0 $\pm$ 0.0       \\
\textbf{N\_bifs}              & 127.0 $\pm$ 77.0   & 49.0 $\pm$ 25.0      & 35.0 $\pm$ 34.0     & 70.0 $\pm$ 180.0    \\
\textbf{N\_branch}            & 260.0 $\pm$ 150.0  & 101.0 $\pm$ 50.0     & 70.0 $\pm$ 68.0     & 140.0 $\pm$ 370.0   \\
\textbf{Width}                & 25.0 $\pm$ 13.0    & 520.0 $\pm$ 140.0    & 225.0 $\pm$ 97.0    & 160.0 $\pm$ 120.0   \\
\textbf{Height}               & 53.0 $\pm$ 32.0    & 690.0 $\pm$ 150.0    & 360.0 $\pm$ 130.0   & 210.0 $\pm$ 130.0   \\
\textbf{Depth}                & 22.0 $\pm$ 14.0    & 0.07 $\pm$ 0.37      & 174.0 $\pm$ 67.0    & 80.0 $\pm$ 52.0     \\
\textbf{Diameter}             & 0.254 $\pm$ 0.003  & 1.0 $\pm$ 0.0        & 1.0 $\pm$ 0.0       & 1.0 $\pm$ 0.0       \\
\textbf{EucDistance}          & 56.0 $\pm$ 25.0    & 610.0 $\pm$ 110.0    & 400.0 $\pm$ 130.0   & 260.0 $\pm$ 130.0   \\
\textbf{PathDistance}         & 109.0 $\pm$ 42.0   & 840.0 $\pm$ 190.0    & 960.0 $\pm$ 330.0   & 550.0 $\pm$ 320.0   \\
\textbf{Branch\_Order}        & 28.0 $\pm$ 10.0    & 13.3 $\pm$ 3.7       & 15.6 $\pm$ 9.8      & 15.5 $\pm$ 9.3      \\
\textbf{Contraction}          & 0.861 $\pm$ 0.017  & 0.947 $\pm$ 0.017    & 0.771 $\pm$ 0.034   & 0.781 $\pm$ 0.04    \\
\textbf{Fragmentation}        & 2350.0 $\pm$ 800.0 & 1160.0 $\pm$ 480.0   & 320.0 $\pm$ 280.0   & 490.0 $\pm$ 770.0   \\
\textbf{Partition\_asymmetry} & 0.661 $\pm$ 0.047  & 0.584 $\pm$ 0.069    & 0.62 $\pm$ 0.11     & 0.584 $\pm$ 0.072   \\
\textbf{Pk\_classic}          & 1.998 $\pm$ 0.014  & 2.0 $\pm$ 0.0        & 2.0 $\pm$ 7.3e-08   & 2.0 $\pm$ 0.0       \\
\textbf{Bif\_ampl\_local}     & 98.5 $\pm$ 5.6     & 92.4 $\pm$ 6.6       & 103.0 $\pm$ 10.0    & 94.1 $\pm$ 9.4      \\
\textbf{Fractal\_Dim}         & 1.128 $\pm$ 0.017  & 1.0129 $\pm$ 0.0064  & 1.103 $\pm$ 0.017   & 1.118 $\pm$ 0.034   \\
\textbf{Bif\_ampl\_remote}    & 93.1 $\pm$ 4.8     & 88.5 $\pm$ 7.6       & 94.3 $\pm$ 9.6      & 90.4 $\pm$ 9.4      \\
\textbf{Length}               & 410.0 $\pm$ 140.0  & 6400.0 $\pm$ 2600.0  & 2800.0 $\pm$ 2400.0 & 2300.0 $\pm$ 3000.0
\end{tabular}
\end{table}

\subsection{Complex networks and modularity}

The \emph{literal modularity}~\cite{da2022literal} has been considered as a possible complement to other related approaches (e.g.~\cite{newman2006modularity}), in order to focus on the relative number of edges within a subgraph candidate to be a module while making no hypothesis on any of the other properties of the subgraph or network to which it belongs. Despite its simplicity, the literal modularity has been found to provide interesting subsidies for the quantification of the separation between subgraphs (or groups) preliminarily identified (supervised recognition) in a network.

Given a network $A$ and any of its subgraphs $S$, the literal modularity of the latter can be calculated as:
\begin{equation}
    \mathcal{L}(S) = \frac{e_i(S)} {e_e(S)}
\end{equation}
where $e_i(S)$ is the total number of directed/undirected edges within the subgraph and $e_e(S)$ is the total number of directed/undirected between the subgraph and the remainder of the network. In the case of $S$ being an isolated connected component, its literal modularity can be calculated by making $e_e(S) =1$, which is the larger possible modularity next to being infinite. A subgraph measure called normalized cut or conductance (e.g.~\cite{Chung:1997,Leskovec}) has been described that is related to the reciprocal of the above quantity $\mathcal{L}(S)$.

Given a network partitioned into $M$ subgraphs $S_k$, $k = 1, 2, \ldots, M$, the overall literal modularity can be quantified in terms of its geometric or arithmetic average, which tends to penalize  distinct individual literal modularities. In the present work, we instead adopt the arithmetic average, in which case the overall literal modularity of a set $\sigma$ of subsets $S_k$ of a given network can be expressed as:
\begin{equation}
    \mathcal{L}(\sigma) = \frac{1}{M} \sum_{k=1}^M \mathcal{L}(S_k).
\end{equation}

In the present work, we optimize the literal modularity in a specific way, namely using a supervised manner, since the categories are available \emph{a priori}. This means that the original categories are taken for granted, providing the reference while quantifying the literal modularity. Thus, by relaxing additional constraints such as uniformly random connections within the modules, the supervised maximization of the literal modularity can proceed with enhanced flexibility while exploring the parametric configurations. Other problems involving the definition of categories, including non-supervised approaches, would need to be addressed by distinct approaches.

In this manner, here, the optimization proceeds not respectively to find a suitable number of modules and their respective content, which is already given, but to determine the few parameters in the coincidence similarity approach so that the literal modularity can be optimized for the provided modules. Though, as discussed in~\cite{da2022literal}, additional requirements on the modules, such as adherence to some specific node distribution, can also be eventually incorporated, the present work aims at finding the coincidence method parameters that optimize the separation of the modules from the remainder of the network as quantified by the literal modularity alone, given that the original identification of the groups in NeuroMorpho.org is taken as reference.

\subsection{Real-valued coincidence similarity}

One of the most frequently performed operations in the physical sciences corresponds to comparing or measuring the difference (distance) or similarity (proximity) between two values or two generic mathematical or physical entities.

While approaches such as the Jaccard (e.g.~\cite{Jaccard1, jaccard1901etude, Samanthula, Jac:wiki, schubert2014note}) and S\o{}rensen-Dice (e.g.~\cite{sorensen1948}) indices are often employed to compare sets and categorical data, the cosine similarity, Pearson correlation coefficient, and Euclidean distance are frequently applied to quantify the similarity/difference between numeric data.

More recently~\cite{da2021multiset,costa2021similarity,da2021further}, the \emph{coincidence similarity index} was introduced as an enhanced version of the Jaccard index that incorporates the quantification of the relative interiority between the compared data as well as a generalization of the Jaccard index to real, possibly negative values. More specifically, the coincidence index can be understood as corresponding to the product of the real-valued Jaccard and interiority (or overlap~\cite{vijaymeena}) indices. As such, the coincidence index can be applied to compare, in a more strict (enhanced selectivity and sensitivity), the similarity between two generic numeric mathematical entities such as functions, vectors and matrices.

In its parameterless version, the real-valued coincidence similarity between two non-zero real-valued vectors $\vec{x}$ and $\vec{y}$ can be simply expressed as:
\begin{align}
    \mathcal{C}_R(\vec{x},\vec{y}) = 
         \mathcal{J}_R(\vec{x},\vec{y}) \ 
     \mathcal{I}_R(\vec{x},\vec{y}) =  \nonumber \\
    = \left( \frac{\sum_{i=1}^N sign(x_i y_i) \min \left\{ \vert x_i\vert, \vert y_i\vert \right\} }  {\sum_{i-1}^N \max \left\{ \vert x_i\vert, \vert y_i\vert  \right\} }      \right)    \left(
    \frac{\sum_{i=1}^N  \min \left\{ \vert x_i\vert, \vert y_i\vert \right\} }  {\min \left\{ S_{\vec{x}},  S_{\vec{y}} \right\} }    \right),
\end{align}
where:
\begin{equation}
    S_{\vec{x}} = \sum_{i-1}^N \vert x_i\vert; \quad  S_{\vec{y}} = \sum_{i-1}^N \vert y_i\vert. 
\end{equation}
It can be verified that $0 \leq \mathcal{C}_R(\vec{x},\vec{y})  \leq 1$. See~\cite{mirkin,akbas,costa2023mulsetions} for related approaches addressing signed similarity in analogy to L1 norms.

A parameterized version of the coincidence index has been described~\cite{costa2021similarity,CostaCCompl,da2021multiset} that incorporates a parameter $\alpha$, with $0 \leq \alpha \leq 1$, that can be used to control the relative contribution of the pairs of values of each feature that have the same or opposite sign. Identical contribution is implemented for $\alpha=0.5$, in which case the parameterized coincidence similarity index becomes identical to its parameterless version. For $ \alpha > 0.5$, the contribution of compared values with the same becomes dominant, implying the network becomes more interconnected. As $\alpha$ is progressively increased, the existing connections are kept, while the strictness of the comparison is relaxed respectively to negative pairwise associations (non-correlated).

The parameterized version of the coincidence similarity index between two real-valued vectors $\vec{x}$ and $\vec{y}$ can be expressed as:
\begin{equation}
  \mathcal{C}_R(\vec{x},\vec{y},\alpha) = 
     \mathcal{J}_R(\vec{x},\vec{y},\alpha) \ 
     \mathcal{I}_R(\vec{x},\vec{y}) 
\end{equation}
where:
\begin{equation}
    J(\vec{x},\vec{y}, \alpha) = \alpha S_p - (1-\alpha)S_n,
\end{equation}

\begin{equation}
    S_p(\vec{x},\vec{y}) = \frac{\sum_i \vert s_{x_i}+s_{y_i}\vert \ min\lbrace\vert x_i\vert,\vert y_i\vert\rbrace}{\sum_i max\lbrace \vert x_i\vert,\vert y_i\vert\rbrace},
\end{equation}

\begin{equation}
    S_n(\vec{x},\vec{y}) = \frac{\sum_i \vert s_{x_i}-s_{y_i}\vert \ min\lbrace\vert x_i\vert,\vert y_i\vert\rbrace}{\sum_i max\lbrace \vert x_i\vert,\vert y_i\vert\rbrace},
\end{equation}
and
\begin{equation}
    \mathcal{I}(\vec{x},\vec{y}) = \frac{\sum_i min\lbrace\vert x_i\vert,\vert y_i\vert\rbrace}{min\lbrace \sum_i\vert x_i\vert,\sum_i\vert y_i\vert\rbrace}.
\end{equation}
The coincidence (as well as the Jaccard) index can be also modified as~\cite{da2021multiset,da2021further}:
\begin{equation}
  \mathcal{C}_R(\vec{x},\vec{y},\alpha,D) = 
    \left[ \mathcal{J}_R(\vec{x},\vec{y},\alpha) \ 
     \mathcal{I}_R(\vec{x},\vec{y}) \right]^D
\end{equation}
with $D$ being a non-negative value (further generalizations are possible but are not considered in the present work for simplicity's sake).

The higher the value of the exponent $D$, the sharper (and more strict) the similarity comparison becomes, implying the network becomes interconnected only by the strongest similarities.

\subsection{The coincidence methodology}\label{sec:coincidence_method}

The strict comparison performed by the coincidence similarity has been employed~\cite{CostaCCompl} as a means to translate datasets, in which each data element is characterized in terms of $M$ measurements, into a respective coincidence network. Each of the data elements is represented as a node, while the coincidence similarity between the features of pairs of data elements defines the weights of the respective links.

In case the feature values have distinct ranges of magnitude, it is often interesting to preliminarily standardize~\cite{gewers2021principal} the values respectively to each adopted feature. This can be done by subtracting the original values from their average and dividing by their respective standard deviation. So normalized, each of the features will have null means and unit standard deviation, with most of the new values resulting within the interval $[-2,2]$. It should be kept in mind that the adoption of specific normalization approaches can significantly impact the results, so they need to be carefully chosen while taking into account the specific requirements of each problem. Our particular choice of standardization is partly taken given that we are interested in quantifying joint variations of the measurements.

Transforming a dataset into a respective network by using the coincidence methodology allows several interesting features including the visualization of the interrelationships between the data elements while taking into account \emph{all} the original features. In addition, the observation of modules or communities (e.g.~\cite{CostaCCompl}) indicates that the data elements are organized into clusters or categories whose elements share several of their properties while being different from the remainder of entries in the same dataset.  In addition, several measurements used to characterized the topology of complex networks (e.g.~\cite{costa2007characterization}) can then be used to further study the obtained coincidence similarity network.

Given that we have three parameters potentially influencing the obtained results in their specific manners, it becomes possible to perform optimization over the parameter space in order to obtain some maximum (or minimum) value of some property of interest regarding the obtained network topology. For instance, this procedure has been applied respectively to maximizing network modularity (e.g.~\cite{CostaCCompl,costa2022similarity,domingues2022city}) or even combinations of properties such as modularity and number of isolated nodes~\cite{dos2022enzyme}.

In the present work, we aim at finding the optimal configurations of the parameters $T$ and $\alpha$, given some specific exponent value $D$, leading to maximum \emph{literal modularity}~\cite{da2022literal} while taking the original categories as references. Given the motivation of studying and relating the morphology of neuronal cells while taking into account their respective topological features, it becomes of particular interest to obtain maximally modular networks for the choice of features. The literal modularity~\cite{da2022literal} is adopted here because, despite its simplicity, it has been found to provide a more direct quantification of the modularity of a given network while considering the original categories. 

\begin{figure}[!ht]
  \centering
    \includegraphics[width=0.99\textwidth]{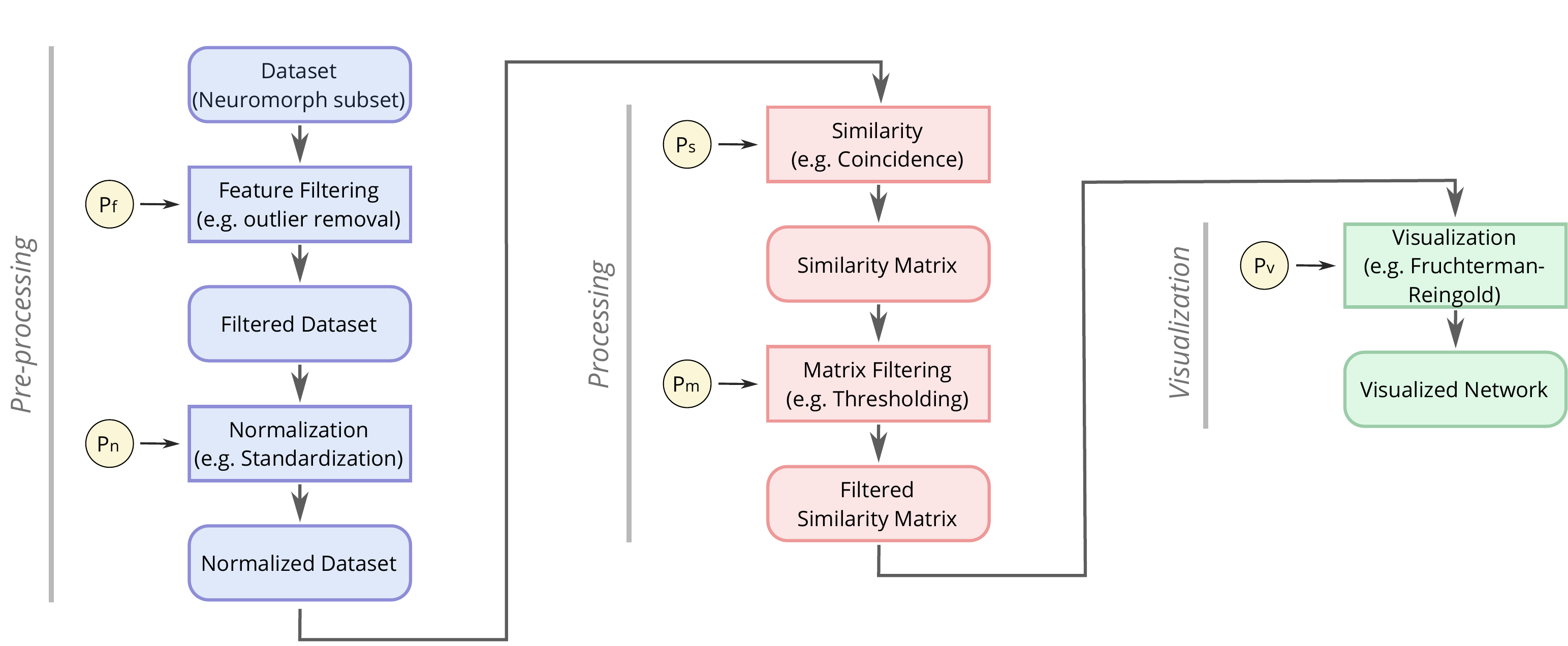}
  \caption {Diagram of the main data (round corner boxes) and methods (rectangles) that constitute the coincidence methodology for representing the neuronal reconstructions, subdivided into three main groups: \emph{pre-processing}, \emph{processing} and \emph{visualization}. The parameters involved in respective methods are shown within yellow disks. }
  \label{fig:diagram}
\end{figure}

Figure~\ref{fig:diagram} summarizes the overall structure --- including data, methods, and parameters --- constituting the coincidence-based approach to translating the neuronal reconstructions into a respective network (following~\cite{CostaCCompl}).  Not shown is the literal modularity optimization in terms of the parameters involved in the \emph{Similarity} and \emph{matrix filtering} stages.

\begin{figure}[!ht]
  \centering
    \includegraphics[width=0.65\textwidth]{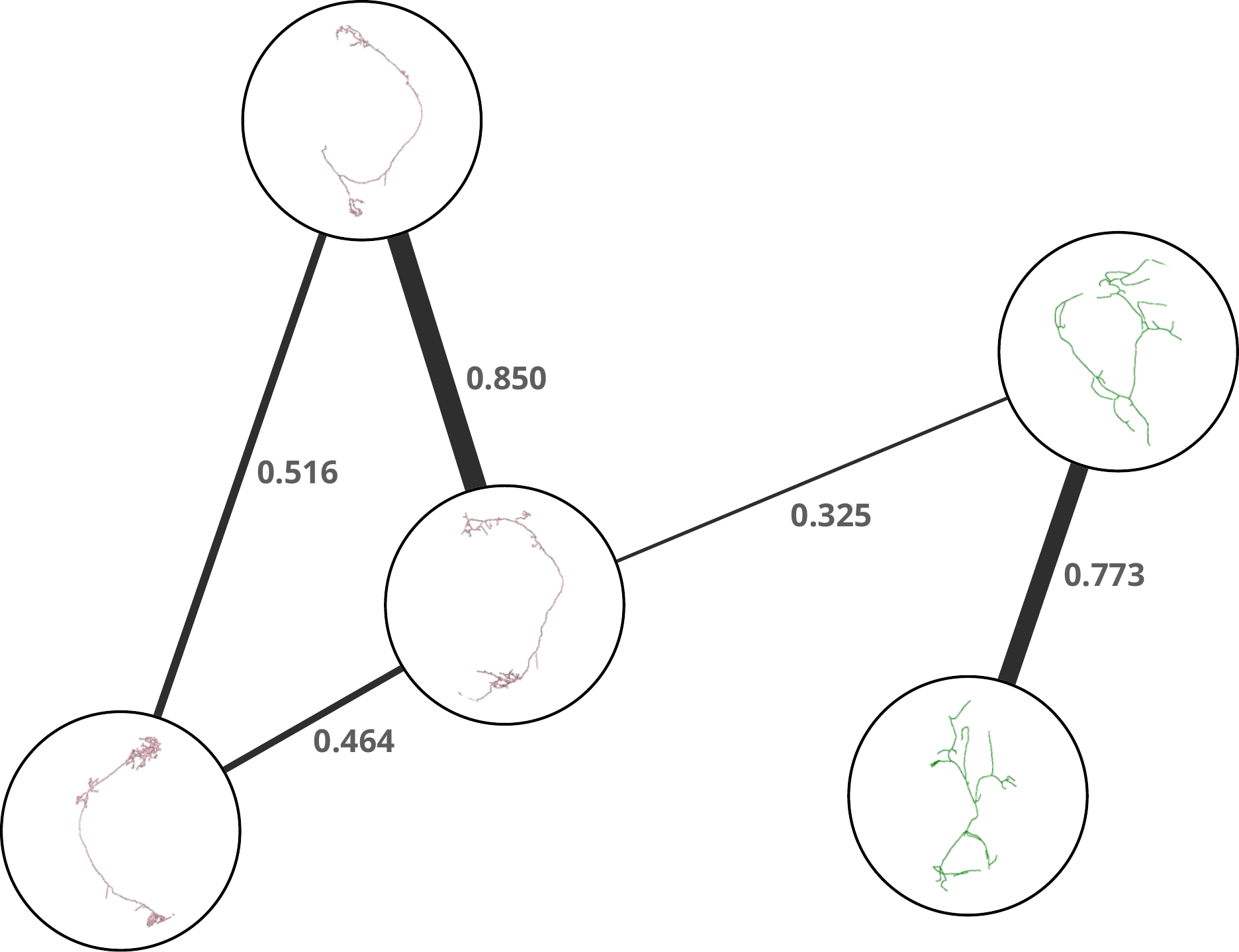}
  \caption {A simple example of the concept underlying the coincidence networks approach respectively to 5 neuronal cells~\cite{lo2015quantification,zheng2018complete} obtained from NeuroMorpho.org, three from the type \emph{Antennal lobe-uniglomerular projection} and two from the type \emph{Chen B-peripheral nervous system-Multidendritic-dendritic arborization}. Each neuron is represented as a network node, and the coincidence similarities between each pair of nodes are reflected as the width of the respective links (values respectively indicated). As adopted henceforth, the coincidence networks are thresholded in order to leave only the strongest relationships, in this case considering the threshold $T= 0.300$.}
  \label{fig:small_net}
\end{figure}

A simple example of the concept of coincidence networks based on the morphological properties of the neuronal cells is depicted in Figure~\ref{fig:small_net}. Each neuron is represented by a respective node, while the coincidence similarity values are reflected as the width (weight) of the respective links.  Therefore, relatively stronger links between two nodes indicate marked similarity (not only morphological but also related to relative sizes) between the respective neurons. In addition, groups of more intensely interconnected nodes (modules or communities) indicate clusters possibly associated with the neuronal types.

\subsection{Study of features influence}

One of the main challenges in artificial intelligence and pattern recognition constitutes the choice of the $M$ features to be adopted for characterizing each of the data elements. The point is that each feature has its intrinsic discriminant effect, which also depends on the specific structure of the available data and the problem being solved, while also being potentially related to other of the adopted features. Importantly, a distinct selection of features, even if differing by only one choice, can strongly impact the resulting analysis and classification.

In the case of the present study, 20 morphological available measurements were taken, each with its intrinsic characteristics and potential. In order to investigate how these features influence the obtained coincidence networks, we applied the procedure described in~\cite{CostaCCompl}, which consists of obtaining a network for each feature combination of interest and then deriving a coincidence similarity network among these resulting structures (see Section~\ref{sec:coincidence_method}). The consideration of modules or communities in the resulting \emph{features network} contributes to a better understanding of the effect of the features on the results.

However, given that 20 features imply in $2^{20}-1$ non-trivial nodes in the respective features network, in the present work we restrict our analysis to configurations involving all the 20 measurements minus one of them so that a features network with $21$ nodes (the configuration with all 20 measurements is also taken into account) is obtained. In this manner, it becomes possible to verify the effect of these feature combinations on the relationships between the networks respectively derived.

Two aspects are of particular importance in this type of analysis of the features influence. First, we have that each obtained community can be understood as a possible explanatory model of the dataset. This is so because the networks in one of these modules will all have similar properties among themselves while differing from the remainder of the networks. Then, the hub (node with the highest strength) of each of these communities can be taken as a possible prototype of that respective model. The larger the number of identified communities, the greater the impact of the features on the modeling can be understood to be. At the same time, communities with large sizes can be understood as being more relevant as potential explanations of the original dataset.

\section{Results and discussion}\label{sec:results}

The 735 cells selected for our analysis initially had their respective features standardized, being subsequently considered in the coincidence methodology. The choice to normalize or not, and which normalization to adopt constitutes an important aspect of pattern recognition, which deserves special attention.  Standardization was implemented in the present work because of the following two reasons: (i) in order to obtain features with more commensurated values variation; and (ii) virtually all measurements, at least in principle, tend to be characterized by joint variations (e.g.~width and length).  

Basically, the value of the coincidence similarity index for each pair of cells was calculated as described in Section~\ref{sec:methods}, yielding a respective $735 \times 735$ matrix with the obtained coincidence values.  

Because the selection of the features to be considered can strongly influence the obtained results in a supervised manner while considering the original categories, we tried not only the configuration involving all measurements but also respective subsets derived from the former by leaving one of the features out in each case. Therefore, a total of $20$ feature combinations were transformed into respective networks with parameter configurations ($D$, $T$, and $\alpha$) that maximize the literal modularity. We tried each of these features combinations respectively to four values of $D$ (namely 1, 2, 4, and 6) and all pairwise combinations between 14 values of $\alpha$ equally spaced between 0.2 and 0.85, and 18 values of $T$ equally spaced between 0.05 and 0.9.

\begin{figure}[!ht]
  \centering
    \includegraphics[width=0.75\textwidth]{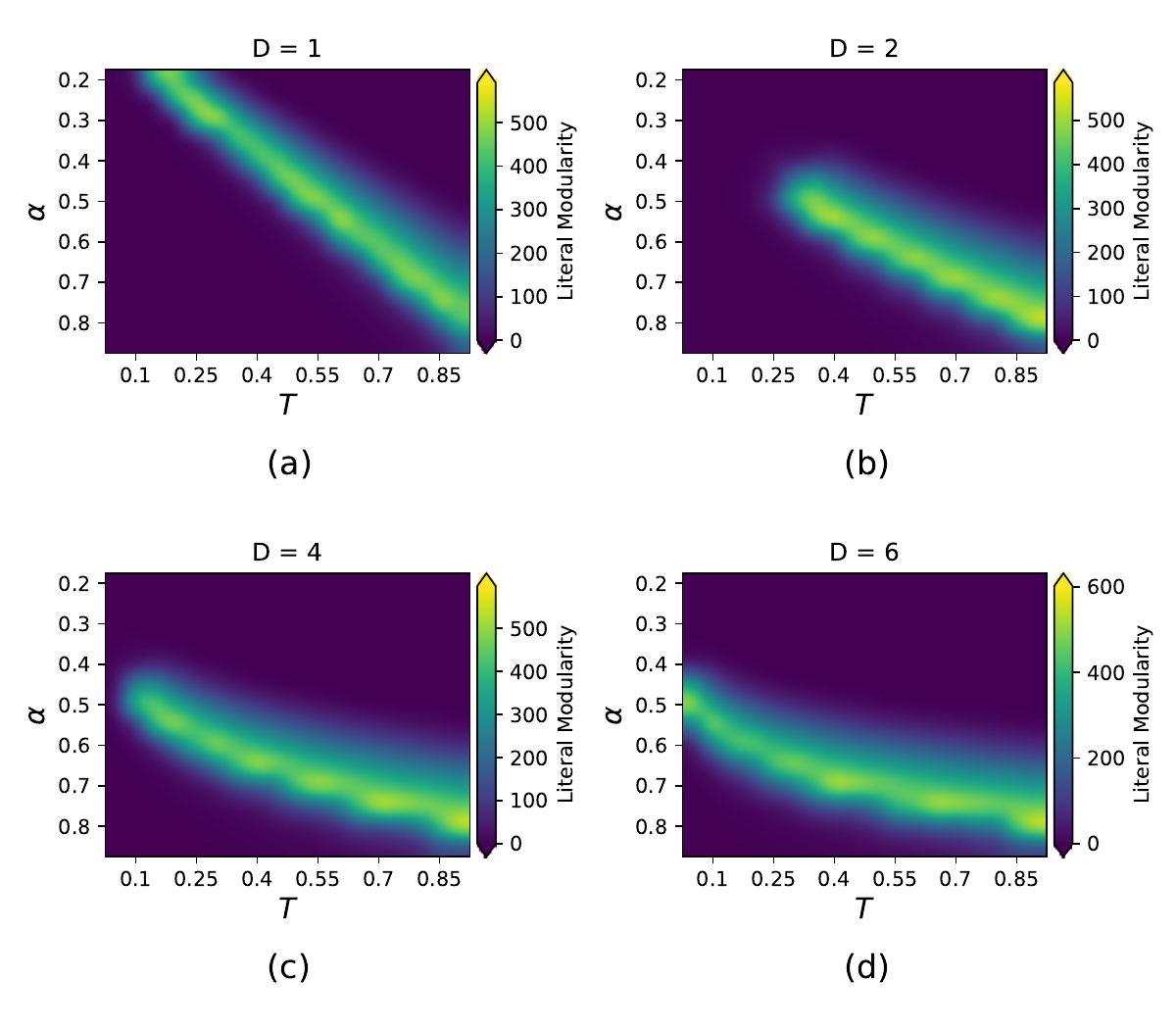}
  \caption {The literal modularities obtained for each of the parameter configurations considered in this work, considering all features. The maximum modularity $L=601.597$ was obtained in a supervised manner (considering the original categories) for $D=6$ respectively to $\alpha = 0.8$ and $T = 0.9$.}
  \label{fig:colormap}
\end{figure}

Figure~\ref{fig:colormap} illustrates the heatmaps of the literal modularity obtained for each of the considered parameter configurations while considering all 20 features, respectively to $D=1$ (a), $D=2$ (b), $D=4$ (c), and $D=6$ (d). The largest values were obtained along a nearly diagonal ridge that changes its shape for different values of $D$. 

\begin{figure}[!ht]
  \centering
    \includegraphics[width=0.7\textwidth]{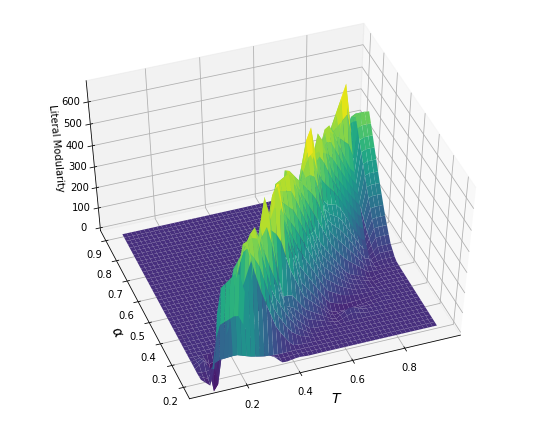}
  \caption {Visualization of the literal modularities, considering the original categories, in terms of $\alpha$ and $T$ as a surface that resembles a ridge.  Interestingly, one of the borders of this ridge is sharper than the other. The largest modularity values are to be found along the peak of the ridge. }
  \label{fig:surface}
\end{figure}

The modularity values obtained for $D=1$ are also shown as a surface in Figure~\ref{fig:surface}, from which the ridge-like appearance of the obtained surface, with one border being more abrupt than the other, can be readily observed. Of particular relevance is the fact that only a small portion of the parameter space, and one that is particularly narrow, yielded large modularity values so that it is important to jointly scan both these parameters~\cite{dos2022enzyme} in case the maximum is to be located.

\begin{figure}[!ht]
  \centering
    \includegraphics[width=0.8\textwidth]{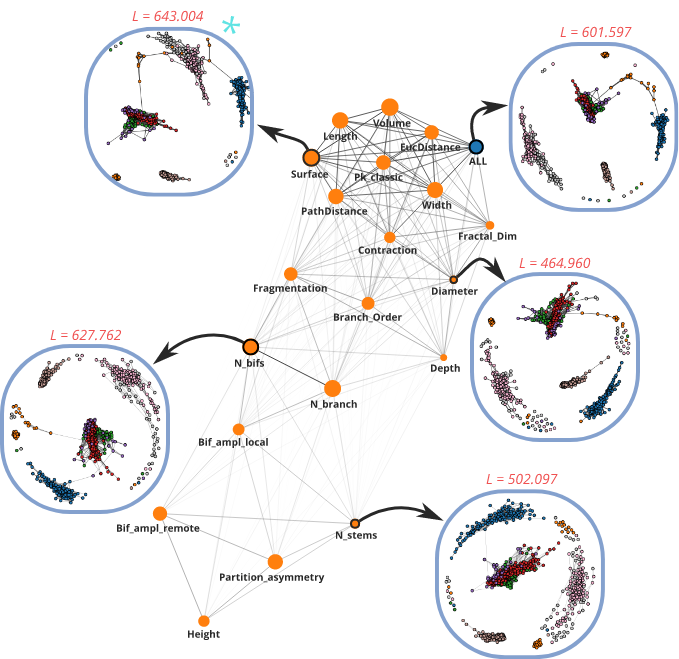}
  \caption {The features network obtained for the selected neuronal reconstructions corresponds to 20 distinct combinations of features. Some respectively obtained maximum literal modularity networks are also shown. It can be observed that the maximum overall literal modularity $L=643.004$ was not obtained for the combination of all 20 features, but for the situation when the measurement \emph{surface} is left out. Also of interest is the concentration of higher modularity values at the top of the network visualization. The cyan asterisk indicates the obtained network with the highest literal modularity.}
  \label{fig:Features}
\end{figure}

Figure~\ref{fig:Features} depicts the features network obtained for the 20 feature combinations considered in this work. Each node corresponds to the modularity-optimal network obtained for the respective feature sets, while the edges indicate the respectively pair-wise similarity. The nodes are labeled respectively to the feature that was respectively excluded from the combination involving all 20 features. The diameter of the nodes is proportional to the respective maximum modularity. Five networks corresponding to respective nodes in this network are also shown for illustrative purposes. The force-directed Fruchterman-Reingold visualization method~\cite{fruchterman1991graph} has been employed in this case, as well as all other visualization in this work.  The feature which, when removed, had the largest impact on the modularity was the \emph{diameter}, which indicates that this measurement has a special contribution to the obtained modularity. Particularly significant impact can also be observed for the features \emph{depth} and \emph{number of stems}.

Another interesting aspect that can be observed from Figure~\ref{fig:Features} concerns the fact that the node corresponding to the maximum literal modularity is more strongly connected to a subgraph that involves the cases \emph{Volume}, \emph{Length}, \emph{EucDistance}, \emph{Pk\_classic}, \emph{Surface}, \emph{PathDistance}, and \emph{Width}.  These nodes, all strongly interconnected, are all characterized by high literal modularity values, indicating that these features complement one another respectively to the characterization of the neuronal types. The obtained network with the highest modularity is indicated by the cyan asterisk, and the network obtained for all features is shown in blue (`ALL').

\begin{figure}[!ht]
  \centering
    \includegraphics[width=0.95\textwidth]{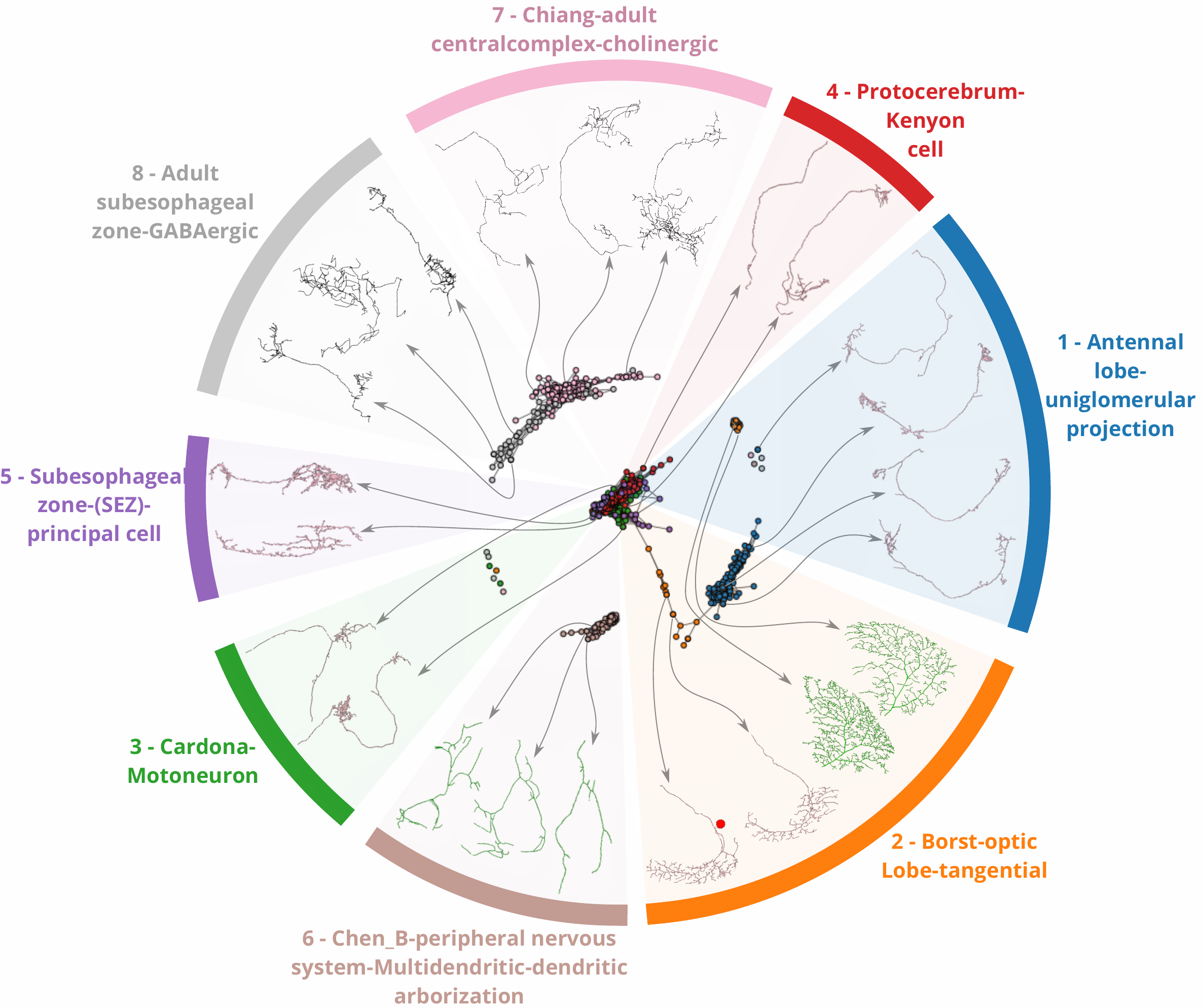}
  \caption {The maximum literal modularity neuromorphic network obtained in this work, presenting the similarity interrelationship between the considered 735 neuronal reconstructions as well as their interconnectivity. A highly modular network has been obtained, with compact clusters corresponding to the original neuronal types. In addition, morphologically similar neurons have been mapped into adjacent modules, while more distinct types are to be found in separated components of the network. Some examples of neurons~\cite{zheng2018complete,boergens2018full,cuntz2013preserving,heckscher2015even,eichler2017complete,schlegel2016synaptic,lo2015quantification,chiang2011three,shih2015connectomics} are illustrated. Or particular interest is the separation of the \emph{Borst-optic Lobe-tangential} neuron type into two distinct groups: one more compact and isolated, and another implementing a bridge between two of the larger obtained connected components.}
  \label{fig:ALL}
\end{figure}

The maximum modularity neuromorphic network is shown in Figure~\ref{fig:ALL}. Each node corresponds to one of the 735 neuronal reconstructions, and the links between pairs of nodes are shown with width proportional to their pair-wise coincidence similarity. Examples of neuronal reconstructions have also been provided respectively to some of the nodes. A number of interesting aspects can be inferred from the obtained network. First, we have that the obtained network is not only highly modular, but its mostly compact modules correspond accurately to the chosen 8 types of neurons. In addition, their morphological relationship is further characterized by the respectively obtained interconnections. Except for the \emph{Borst-optic Lobe-tangential} neuron type, which has been split into two separated groups, neuronal types that are morphologically similar were mapped into adjacent modules, while more distinct reconstructions were associated with well-separated modules. Interestingly, one of the groups into which the \emph{Borst-optic Lobe-tangential} neuron type has been partitioned can be observed to implement a bridge between two of the larger modules.
Also worth noticing is the fact that this group yielded a separated network, probably as a consequence of the literal modularity not imposing additional restrictions such as uniformly random connections within the modules, therefore allowing increased flexibility while searching, in supervised manner, for the maximum value of the literal modularity.

\begin{figure}[!ht]
  \centering
    \includegraphics[width=.80\textwidth]{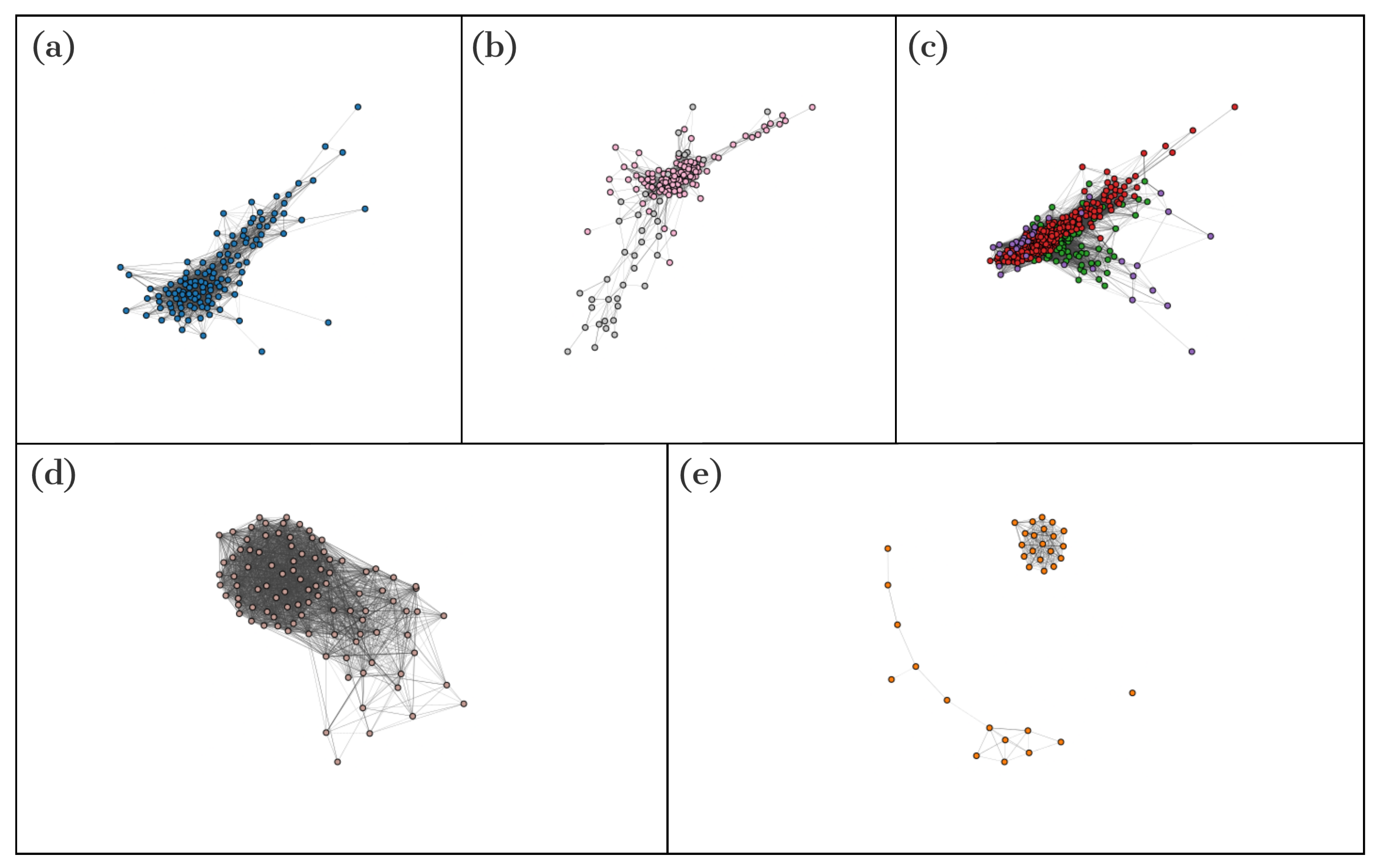}
  \caption {Separated visualization of the five main detected groups, providing a better identification of the interconnections between the respective neurons, each presenting a specific organization and interconnecting structures between the respective individual neuronal cells.}
  \label{fig:singular_groups}
\end{figure}

Figure~\ref{fig:singular_groups} illustrates the five largest modules in the modularly optimal obtained network, so as to allow better visualization of the interconnectivity within each module. Several interesting characteristics can be observed. Except for group (c), all other groups are uniform, i.e.~contain neurons from the same type. Groups (a), (b), and (d) are characterized by a denser core surrounded by respectively less similar neurons. These cores can then be understood as being composed of neurons that are markedly similar to one another. Group (e) presents an almost completely interconnected group, as well as a with a sparse component. Though group (c) contains neurons from three neuronal types, most of the respective cells are closely interconnected, suggesting that most of the cells in this group are strongly similar. Also respectively to this group, it can be observed that the group shown in red appears between the other two groups, suggesting that the neurons in this group have intermediate morphological characteristics.

\begin{figure}[!ht]
  \centering
    \includegraphics[width=.85\textwidth]{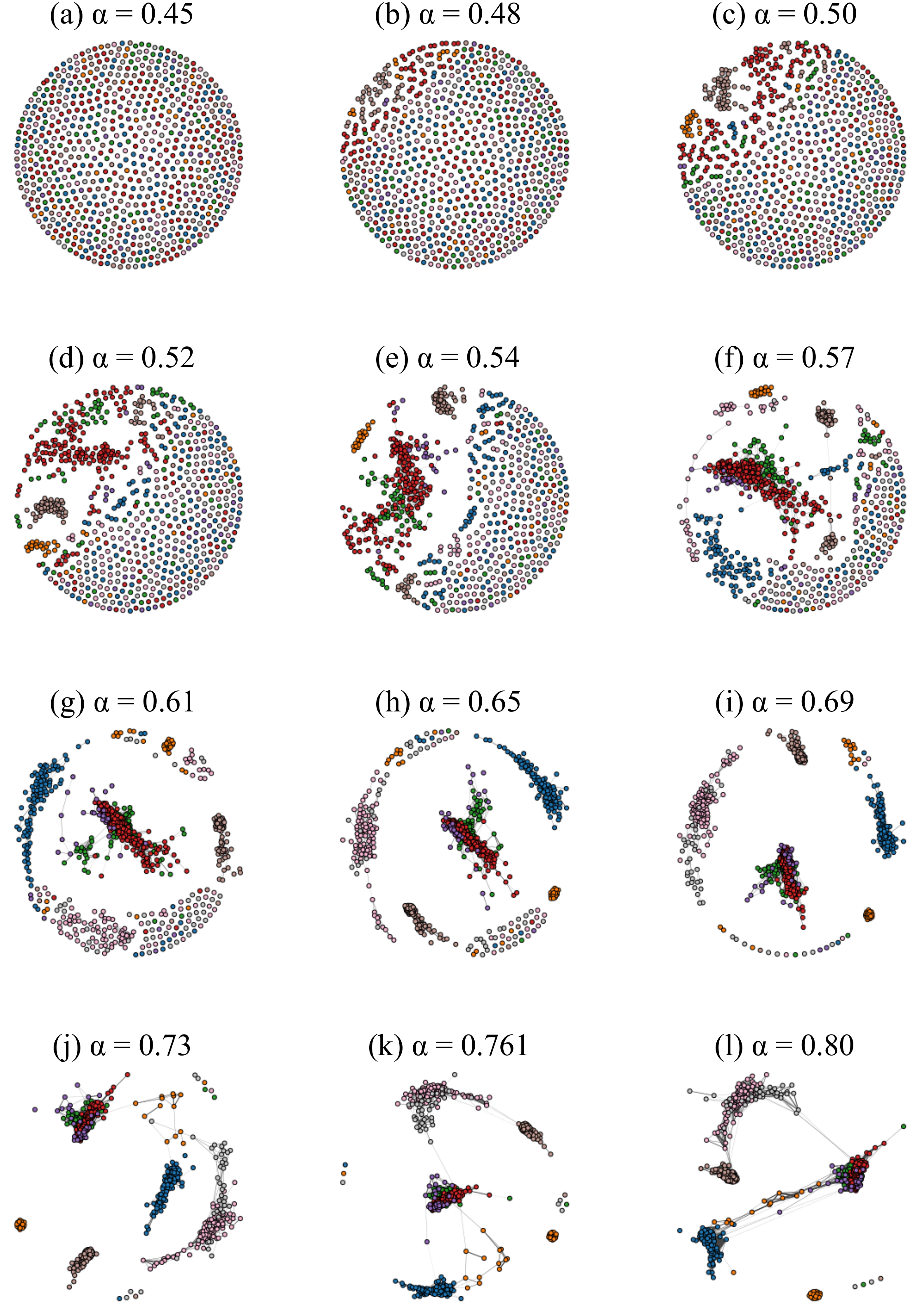}
  \caption {Several coincidence neuromorphic networks obtained for increasing values of $\alpha$, were chosen in order to better illustrate the most relevant mergings between the modules associated with the neuronal types. As $\alpha$ increases, all previous links are maintained while new connections are progressively established in decreasing order of similarity. This type of representation provides a rich indication about how the successive interconnections are established between the neuronal cells, giving rise to respective modules, which are then followed by the progressive merging between those modules.}
  \label{fig:alpha_progress}
\end{figure}

While the coincidence similarity network in Figure~\ref{fig:ALL} corresponds to the parameter configuration for which the literal modularity takes its maximum value, additional valuable information about the interrelationship between the several involved neuronal types and cells can be immediately obtained from the adopted methodology by presenting coincidence networks respectively to several increasing values of the parameter $\alpha$, as depicted in Figure~\ref{fig:alpha_progress}. 

The succession of mergings between neuronal cells, obtained as the parameter $\alpha$ is increased, can be clearly discerned, leading to modules that are mostly uniform regarding the cell types. Initiating with all cells being disconnected (a), we can then identify the brown neurons coalescing. As $\alpha$ increases, we can observe in (c) a larger module of brown cells, as well as newly started groups of red and orange cells. These groups then continue to grow by incorporating new neuronal cells that are mostly from the same neuronal type. In (f) we can perceive the initial stage in the formation of the group of blue cells, as well as a separated group of green cells. The group of pink cells starts coalescing at (g) and then grows successively as the separation between the modules also increases significantly until reaching the maximum modularity at $\alpha=0.8$. Also clearly indicated in this figure are the successive mergings between the groups, which take place mostly from (g) to (l). Observe that, in all cases, once a group starts coalescing, it then becomes more and more compact.

The succession of connections between groups provides a valuable indication of which groups are more similar to one another. For instance, we observe from (i) to (j) the merging of the blue group with the orange group, while from (j) to (k) the brown group is at last linked to the other groups. Interestingly, the compact portion of the orange group remains disconnected until (l) because these groups are almost completely interconnected. It can also be observed from Figure~\ref{fig:ALL} that the maximum identified literal modularity succeeded in identifying the network in which the modules are most well-separated, at least from the visual point of view.

\begin{figure}[!ht]
  \centering
    \includegraphics[width=.25\textwidth]{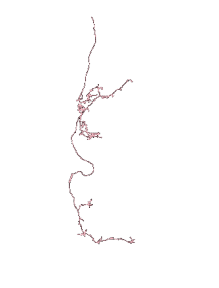} 
    \includegraphics[width=.25\textwidth]{728_out.png} 
    \includegraphics[width=.25\textwidth]{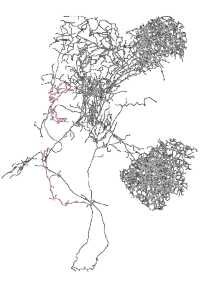} 
  \caption {Three neuronal cells~\cite{takagi2017divergent,shih2015connectomics} remain disconnected even for $\alpha=0.80$, and therefore can be understood as \emph{outliers}. Cell (a) is substantially thicker than other similar cells, with a total volume of $162.739$. The cells in (b) and (c) are also markedly distinct from the other considered cells in both shape and measurements. Though they are similar to one another, about half of their measurements, especially those related to the number of branches, are substantially distinct.}
  \label{fig:outliers}
\end{figure}

Given that three individual neuronal cells remain disconnected even at $\alpha=0.80$, they are here understood as corresponding to \emph{outliers} respectively to the other considered neurons. These three neurons are shown in Figure~\ref{fig:outliers}. In the case of the cell (a), though its shape is similar to the many types of considered neurons and is particular to group 3, which corresponds to its type, it is substantially longer and much thicker than the cells in the other groups. The cells (b) and (c) are both distinct from the other types of neurons but also present markedly different measurements despite their similar shape.

\section{Individual analysis}

In this section, we employ visualization resources (e.g.~\cite{da2022Supervised})
as a complementary means to better understand the several results and findings described previously. This method involves representing each of the types of neurons as a respective square or rectangle containing as many samples as possible (a few of these samples may need to be left out in order to allow adequate factorization of the rectangle sides). Each of the samples within each of the rectangles is represented in terms of their respective features organized and factorized into a respective square and presented in increasing order from left to right and top to down. For better overall contrast in the visualizations, the outliers have been removed. The identified outliers include those shown in Figure~\ref{fig:outliers}.

\begin{figure}[!ht]
  \centering
    \includegraphics[width=.85\textwidth]{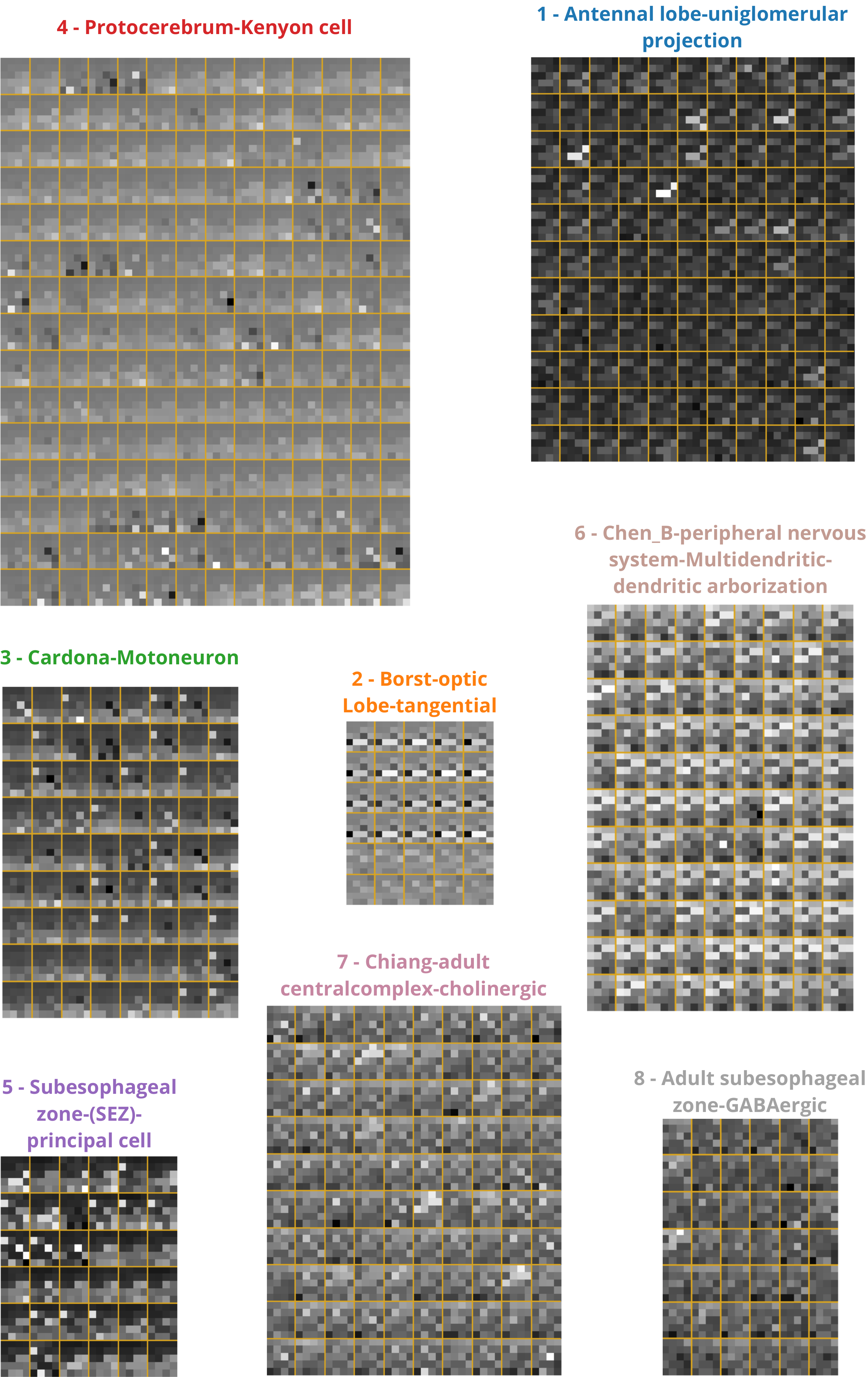}
  \caption{Visualization of the measurements of the complete dataset by representing the cells in each of the eight neuronal groups as rectangles (e.g.~\cite{da2022Supervised}), shown as being delimited by yellow borders. All these rectangles have $5 \times 4$ respective features. The obtained visualization allows several important related aspects to be identified (see text for discussion).}
  \label{fig:vis}
\end{figure}

The result of the application of this visualization approach to the data considered in the present work is presented in Figure~\ref{fig:vis}. Here, we have eight rectangles corresponding to each of the considered neuronal types, while the 20 features are organized into smaller rectangles of size $5 \times 4$, which are shown delimited by the yellow borders. Standardized versions of the original features were considered in Figure~\ref{fig:vis}.

This type of representation provides visual information about the aspects of the original dataset. First, we can see the variations in the number of samples per group. Then, by inspecting inside each group, it is possible to have indications about the relative uniformity between the samples. For instance, types 6 and 7 tend to have more uniform features, while types 5 and 8 are particularly heterogeneous, which implies that these two groups remain strongly compact along the $\alpha$ variation analysis.

Another important effect that can be immediately identified in the visualization in Figure~\ref{fig:vis} concerns the potentially critical influence of some samples in biasing the respective standardization. This can be readily observed in the case of neuronal types 1 and 4, where the presence of a few outliers implied a respective significant scaling/shifting of the values. 

Additional insights can also be obtained regarding the potential clustering between the groups.  For instance, the sample values within the samples in groups 3 and 4 are visually substantially similar, which is consistent with these two groups merging soon along the $\alpha$ variation analysis. Also visually noticeable is the intrinsic similarity between the features of types 7 and 8 of neurons, which was duly reflected in the fact that these two groups tended to merge from the smaller values of $\alpha$.

Therefore, the visualization shown in  Figure~\ref{fig:vis} can be understood as being mostly compatible with the analysis results reported previously in this work. It is important to keep in mind that this type of visual investigation, given its qualitative nature, should by no means be taken as a validation of the obtained results. However, this analysis can have important implications in providing insights about possible \emph{problems} in the approach, as indicated by eventual inconsistencies, discrepancies, instabilities, or other artifacts and effects that may have not been respectively taken into consideration. This type of visualization can also be valuable as a means to preliminary getting acquainted with the general characteristics, possibly contributing to the planning of several aspects of the analysis.

\begin{figure}[!ht]
  \centering
    \includegraphics[width=.85\textwidth]{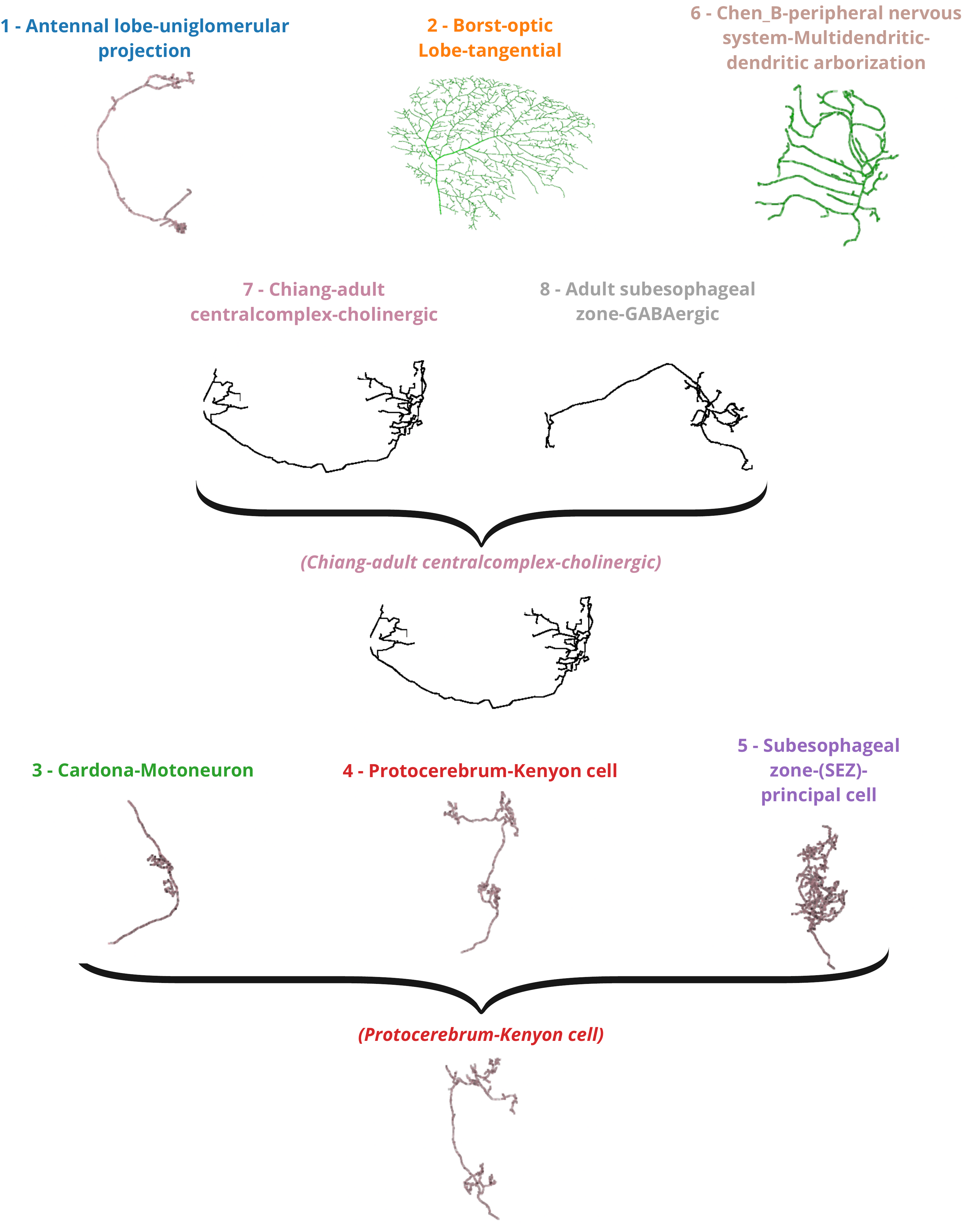}
  \caption{The prototypes of each of the eight original categories~\cite{zheng2018complete,cuntz2013preserving,lo2015quantification,chiang2011three,shih2015connectomics,heckscher2015even,eichler2017complete,schlegel2016synaptic}, as well as of the identified groups.}
  \label{fig:protots}
\end{figure}

To conclude this section, we present in Figure~\ref{fig:protots} the \emph{prototypes}~\cite{costa2022similarity} of each of the eight considered neuronal types. Recall that each prototype corresponds to a neuron associated with the node with the largest strength (i.e.~sum of the similarity values associated with the links of a node) in the respective group. For each neuronal type, the neuron that has the largest sum of coincidence similarities to other nodes is taken as the respective prototype~\cite{CostaSelfCoincidence}. The prototype of the identified groups is the same in every case except for the three groups 7/8 and 3/4/5, so we also include the prototype for these merged identified groups, which turned out to be of type 7 and 4, respectively.

\section{Concluding remarks} \label{sec:conc}

The reconstruction, representation, and characterization of neuronal cells constitutes an important activity as it can provide valuable resources not only for better understanding neuronal shape but also as subsidies for inferring how interrelated these morphological types can be. In the present work, we applied a similarity-based approach to translating a set of neuronal reconstructions, characterized in terms of 20 morphological features, into maximally modular respective networks respectively to the involved parameters. Well-separated groups were obtained that closely reflect the original types of neurons, with the overall interconnectivity providing a rich and comprehensive characterization of the similarity relationship between the 735 considered neuronal reconstructions. The effect of several feature combinations was also investigated in terms of a respective features network, which indicated the impact of individual features on the achieved modularity, which was found to be mostly comparable, though three features (\emph{diameter}, \emph{depth} and \emph{number of stems}) had a more substantial impact on the obtained modularity.

The interconnectivity within each of the obtained separated components was also illustrated in terms of isolated coincidence similarity networks, allowing the identification of distinct types of morphological relationships at the smaller scales of similarity, where enhanced accuracy is often desirable. The ability of the coincidence method to reveal the interaction between the identified modules in terms of progressively smaller similarity values was also applied, by considering successively higher values of the parameter $\alpha$, providing a comprehensive indication of similarity relationships between the neuronal cells from the smallest to the largest scales, allowing the identification of the sequence of merging between cells and then between groups. The potential of the reported methodology for identifying and characterizing possible outlier cells was also illustrated respectively to three neurons that remained isolated even at the higher considered values of $\alpha$.

It should be kept in mind that the described results are specific to the considered dataset, hypotheses, and main questions addressed here. Distinct methodological choices and parameter configurations apply to other datasets and applications. For instance, the literal modularity adopted in the present work has been mainly allowed by the fact that the respective cell categories were already available (supervised approach). Other types of modularity will be required otherwise. 

Several possible developments of the method presented in this work are possible. In particular, it would be important to further investigate and validate the described approaches respectively to a wider range of datasets, methodological choices, parametric configurations, and requirements specific to other problems and applications. For instance, other types of cells could be analyzed and characterized. The proposed methodology can also be directly adapted to the analysis of other neurological structures at diverse scales, such as glial cells, parts of a neuron (e.g.~dendrites or axons~\cite{Cervantes}), as well as the shape and other properties of cortical regions.

\section*{Acknowledgments}
Alexandre Benatti thanks Coordenação de Aperfeiçoamento de Pessoal de Nível Superior - Brasil (CAPES) - Finance Code 001. Henrique F. de Arruda acknowledges FAPESP for sponsorship (2018/10489-0, from 1st February 2019 until 31st May 2021). Luciano da F. Costa thanks CNPq (grant no. 307085/2018-0) and FAPESP (grant 15/22308-2).

\bibliography{ref}
\bibliographystyle{unsrt}

\end{document}